\documentclass{elsart}
\usepackage{amssymb,amsmath}
\usepackage{psfig}
\begin{document}
\begin{frontmatter}

\title{Kinetic energy functional for  Fermi vapors in spherical
harmonic confinement}   
\author[sns]{A. Minguzzi,}
\author[oxf,ruca] {N.~H. March} and
\author[sns]{M.~P. Tosi\corauthref{cor1}} 
\corauth[cor1]{Corresponding author.}

\address[sns]{INFM and
Classe di Scienze, Scuola Normale Superiore, 
56126 Pisa, Italy}
\address[oxf] {Oxford University, Oxford, England}
\address[ruca] {Department of Physics, University of Antwerp (RUCA),
Antwerp, Belgium}

\maketitle                
\begin{abstract}
Two equations are constructed which reflect, for fermions moving
independently in a spherical  harmonic 
potential, a differential  virial theorem and a
relation between the turning points of kinetic energy and particle
densities. These equations are  used to derive 
a differential  equation for the particle
density and a non-local kinetic energy functional.
\end{abstract}

\end{frontmatter}

\section{Introduction}
There is currently considerable experimental activity in the area of
harmonically confined fermion vapors \cite{jila_exp,paris_exp}. This
has been the motivation for the present Letter, in which equations are
set up involving the particle density $\rho(r)$ and the
kinetic energy 
density $t(r)$. We have written both quantities solely as functions of
the radial distance $r$, as we shall be dealing with an arbitrary
number, say $(M+1)$ closed shells in an isotropic external potential. 

Two equations form the basis for what constitutes a density functional
theory of independent harmonically confined fermions in three
dimensions.  The first is a form of differential virial equation,
specific to an isotropic oscillator potential
\begin{equation}
V(r)=\frac{1}{2} m\omega^2 r^2
\label{potential}
\end{equation}
where $\omega^2=k/m$, with $k$ the force constant and $m$ the fermion
mass. The second equation relates the turning points of the particle
density $\rho(r)$ and kinetic energy density $t(r)$, which we define
from the wave function form
\begin{equation}
t(r)=-\frac{\hbar^2}{2m} \sum_i \psi_i(r) \nabla^2 \psi_i(r)\;.
\label{ti}
\end{equation}
Two other forms of the kinetic energy density will be used below, the
first from the gradient of the wave function 
\begin{equation}
t_G(r)=\frac{\hbar^2}{2m} \sum_i |\nabla\psi_i(r)|^2
\label{tigi} 
\end{equation}
and the other derived as the average of $t(r)$ and $t_G(r)$,
\begin{equation}
\bar t(r)=\frac{t(r)+t_G(r)}{2}\;.
\label{tibar}
\end{equation}

\section{Differential virial equation}
As to the differential virial equation mentioned above, the early work
of March and Young \cite{march_young}, which however was restricted to
one-dimensional motion 
along the $x$ axis, say, gave by expansion of
the equation of motion of the Dirac density matrix around its
diagonal:
\begin{equation}
\frac{ d t(x)}{d x}= -\frac{1}{2}\rho(x) \frac{d V(x)}{d
x}-\frac{\hbar^2}{8m} \rho'''(x).
\label{vir1d}
\end{equation}
This was termed a differential virial equation since, by
multiplying by $x$ and integrating over all $x$ the usual integral
virial theorem was recovered. We note from the
definitions~(\ref{ti})-(\ref{tibar}) above, with the consequence that 
\begin{equation}
t_G({\mathbf r})=t({\mathbf r}) +\frac{\hbar^2}{4m}\nabla^2
\rho({\mathbf r})\;,
\label{tigivsti}
\end{equation}
that Eq.~(\ref{vir1d}) can be rewritten 
\begin{equation}
\frac{d \bar t(x)}{dx}=-\frac{1}{2}\rho(x) \frac{dV(x)}{dx}\;.
\label{virmod1d}
\end{equation}

While, for one dimension Eq.~(\ref{virmod1d}) is valid for a general
confining potential $V(x)$, its generalization of present interest refers 
to a three-dimensional isotropic harmonic oscillator with closed-shells
occupancy.
It is not our purpose in this Letter to attempt any general
proof, but we shall mention later a number of convincing confirmations
of our result. This is 
\begin{equation}
\frac{\partial \bar t(r)}{\partial r}=-\frac{3}{2} \rho(r)
\frac{\partial V(r)}{\partial r}\;,
\label{vir3d}
\end{equation} 
but with $V(r)$ now restricted to the harmonic form~(\ref{potential}).
The factor 3/2, by comparison with Eq.~(\ref{virmod1d}), is readily
understood as due to dimensionality, as is confirmed by multiplying
Eq.~(\ref{vir3d}) throughout by $r$ to form the virial of the force
$-\partial V(r)/\partial r$. A volume integration, followed by
integration by parts on the LHS of the resulting equation yields
\begin{equation}
2 T =\int d^3r  \,\rho(r) r \frac{\partial V(r)}{\partial r}
\label{virint}
\end{equation}
where $T=\int d^3r\, \bar t(r)$ is the total kinetic
energy.

\begin{figure}
\centerline{\psfig{file=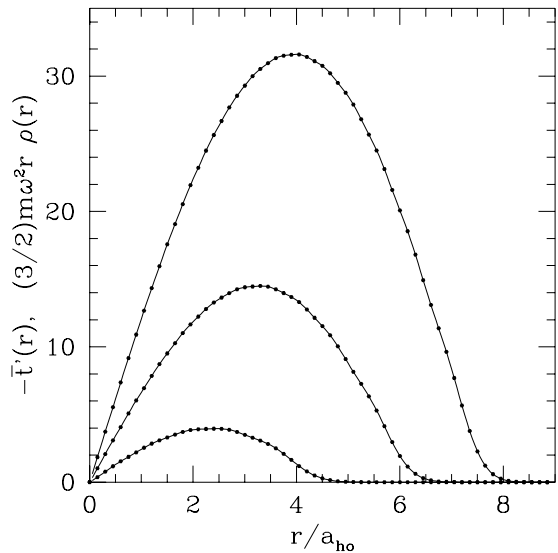,width=0.55\linewidth}}
\caption{Test of the differential virial equation: negative of
 $\partial \bar t(r)/\partial r$ (solid line) and $(3/2) m\omega^2 r
 \rho(r)$ (dots) in units $\hbar \omega/a_{ho}^4$
 as functions of the radial coordinate $r/a_{ho}$, for a 3D Fermi
gas in isotropic harmonic confinement  with 10, 20 and 30  filled
 shells (from bottom to top).}
\label{fig1}
\end{figure}

It is a very
simple matter to verify Eq.~(\ref{vir3d}) directly for the lowest
shell only occupied, when  $\psi_{M=0}(r)={A}\exp(-r^2/2a_{ho}^2)$
and $\rho_{M=0}(r)={A}^2 \exp(-r^2/a_{ho}^2)$ with
$a_{ho}=\sqrt{\hbar/m\omega}$ being the harmonic oscillator
length and $A$ a normalization constant. 
More generally, numerical tests of Eq.~(\ref{vir3d}) are shown
in Fig.~\ref{fig1} for three values of the number of filled
shells. In these calculations we have used the explicit expressions
for the particle and kinetic energy densities reported by Brack and
van Zyl \cite{brack} in terms of Laguerre polynomials.

As a last remark we note that, although the focus of the present work is
on spherical harmonic confinement, any additional term that may enter
the RHS of Eq.~(\ref{vir3d}) for more general central  $V(r)$
potentials does not contribute to the integral virial. Indeed
Eq.~(\ref{virint}) is the correct form of the virial theorem for such
confinements.

\section{Turning points of particle and kinetic energy densities}
Having given these arguments for the form~(\ref{vir3d}) of the
differential virial equation for spherical  harmonic
confinement, we pass to the second central result of the present study,
namely a relation between the turning points of kinetic energy and
particle densities. Again, we can make a useful analogy with the
one-dimensional harmonic oscillator. With $N$ singly occupied shells
we can write, following Lawes and March \cite{lawes}:
\begin{equation}
\frac{t'(x)}{\rho'(x)}=N\hbar \omega -\frac{1}{2} m \omega^2 x^2\;,
\label{turning1d}
\end{equation}
the single lowest occupancy corresponding to $N=1$.

\begin{figure}
\centerline{\psfig{file=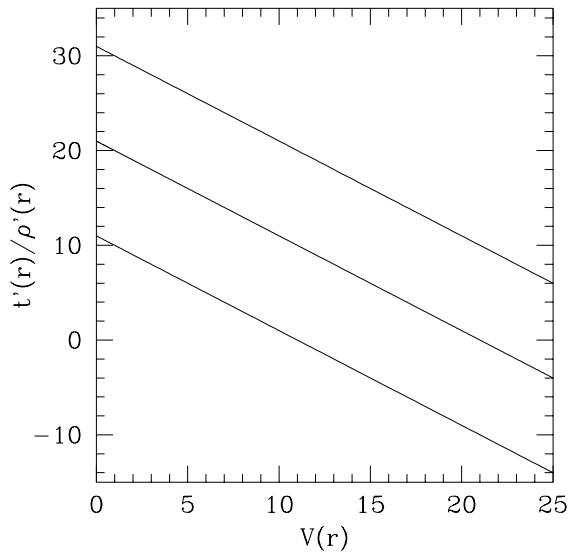,width=0.55\linewidth}}
\caption{Test of the relation between the turning points of particle
and kinetic energy densities:
 $t'(r)/\rho'(r)$  in units $\hbar \omega$
 as functions of the confining potential  $V(r)=m\omega^2 r^2/2$,
for a 3D Fermi gas in isotropic harmonic confinement  with 10, 20 and
 30  filled  shells (from bottom to top).}
\label{fig2}
\end{figure}

We now assert that, for the three-dimensional oscillator potential
energy~(\ref{potential}) the appropriate generalization of
Eq.~(\ref{turning1d}) reads
\begin{equation}
\frac{t'(r)}{\rho'(r)}=(M+2)\hbar \omega -\frac{1}{2}m\omega^2 r^2 \;.
\label{turning}
\end{equation}
Again, no general proof will be attempted. But for the lowest shell
$M=0$ only occupied, we can again use the explicit Gaussian
wave function $\psi_{M=0}(r)$ given above to calculate $t(r)$ in
Eq.~(\ref{turning}) and $\rho_{M=0}'(r)$ is also readily
obtained. Equation~(\ref{turning}) is confirmed for $M=0$ after a
short calculation. More generally, Fig.~\ref{fig2} reports numerical
tests of Eq.~(\ref{turning}) for three values of the number of filled
shells. Again the specific expressions of Brack and van Zyl
\cite{brack} for $t(r)$ and $\rho(r)$ have been used.

Further confirmation, now in integral form but for general $M$, comes
by multiplying Eq.~(\ref{turning}) throughout by $r \rho'(r)$ 
and doing a volume integration. Integration by parts
readily yields, invoking the equipartition theorem for the harmonic
oscillator, 
\begin{equation}
T=-(M+2)\hbar \omega \int_0^\infty r\rho'(r) 4\pi r^2 dr\;. 
\end{equation}
But the integral on the RHS is related  to the total number $N$
of fermions occupying $(M+1)$ closed shells, the simple result 
\begin{equation}
\frac{T}{N}=\frac{3}{8} (M+2)\hbar \omega
\end{equation}
following as a consequence of Eq.~(\ref{turning}). This is readily
confirmed by calculating the eigenvalue sum $E$ from the oscillator
levels plus the known degeneracies. Again, for $N=1$ or equivalently
$M=0$, $T=3\hbar \omega/4$ follows immediately from the virial result
$T=E/2$ since $E=3\hbar\omega/2$ for the lowest shell.

Before considering the consequences of Eq.~(\ref{turning}), when
combined with the differential virial theorem, we want to stress that,
for $M>0$, the Pauli Exclusion Principle plays a role in determining
both particle and kinetic energy densities. Therefore, as a further
confirmation of Eq.~(\ref{turning}) we have used explicit harmonic
oscillator wave functions (see {\it e.g.} Morse and
Feshbach \cite{morse_feschbach}) to construct the Dirac density matrix
$\gamma (\mathbf {r,r'})$ for two shells occupied, {\it i.e.} for
$M=1$. Since the degeneracy of the  levels for a given total quantum
number $n$ is $(n+1)(n+2)/2$, one is then dealing for singly occupied
levels with 4 particles, and the Dirac matrix is readily constructed as 
\begin{equation}
\gamma_{M=1}(\mathbf{
r,r'})=(\pi^{1/2}a_{ho})^{-3} [1+2(xx'+yy'+zz')/a_{ho}^2]
\exp[-(r^2+{r'}^2)/2a_{ho}^2]\;.    
\end{equation}
Forming the kinetic energy density from
\begin{equation}
t_{M=1}(r)=-\frac{\hbar^2}{2m}\left.\nabla^2_{\mathbf r'}\gamma_{M=1}(\mathbf{
r},\mathbf{r'})\right|_{\mathbf{ r'}=\mathbf{r}}
\end{equation}
and using the explicit form of $\rho'(r)$ following from
differentiating with respect to $r$ the diagonal density
$\rho_{M=1}(r)=\gamma_{M=1}(\mathbf {r},\mathbf {r'})|_{\mathbf
{r'}=\mathbf{r}}$, 
once again confirms the correctness of Eq.~(\ref{turning}) for
$t'(r)/\rho'(r)$. 
  
\section{Density functional theory}
Given therefore Eqs.~(\ref{vir3d}) and (\ref{turning}) as the two
pillars of the present study we go on to construct the density
functional theory of this problem of three-dimensional harmonic
confinement. First of all, let us use Eqs.~(\ref{tibar})
and~(\ref{tigivsti}) to remove $\bar t(r)$ in Eq.~(\ref{vir3d}) in
favour of $t(r)$ and $\nabla^2\rho(r)$. Then invoking
Eq.~(\ref{turning})  the derivative $\partial
t(r)/\partial r$ can be eliminated, the result being a differential
equation for the particle density $\rho(r)$. This reads
\begin{equation}
\frac{\hbar^2}{8m} \frac{\partial}{\partial r} \left[\nabla^2
\rho(r)\right] +[(M+2)\hbar \omega -V(r)]\rho'(r)
+\frac{3}{2}\frac{\partial V(r)}{\partial r} \rho(r)=0\;,
\label{eqdiff}
\end{equation}
which we must again emphasize is valid for the three-dimensional
potential energy~(\ref{potential}). Equation~(\ref{eqdiff}) is evidently a
third-order, linear, homogeneous equation for the particle density $\rho(r)$
of $(M+1)$ closed shells. Since the pioneering work of von
Weizs\"acker \cite{weis}, it has been clear that a major objective of
density functional theory is to construct directly the fermion
particle density from a given one-body potential. Equation~(\ref{eqdiff})
achieves that aim quite explicitly for closed shells in the harmonic
potential~(\ref{potential}). It is the three-dimensional
generalization of the early result of Lawes and March \cite{lawes} for
one-dimensional harmonic confinement.

But there is a further consequence of the two basic
equations~(\ref{vir3d}) and (\ref{turning}). For
one can alternatively eliminate the potential, in order to directly
relate kinetic energy and particle densities. One finds almost
immediately 
\begin{equation}
\frac{\partial \bar t(r)}{\partial r}=\frac{3}{2} \rho(r)
\frac{\partial}{\partial r}\left[\frac{t'(r)}{\rho'(r)}\right]\;.
\end{equation}
As emphasized earlier, $\bar t(r)$ and $t(r)$ differ only by a constant
times the Laplacian of the particle density. The resulting relation
can be integrated, using an appropriate integrating factor, to give
$t(r)$ in terms of integrals of $\rho(r)$ and its low-order
derivatives. Explicitly we find 
\begin{equation}
\frac{\partial}{\partial r}
\left[\frac{t'(r)}{\rho'(r)\rho^{2/3}(r)}\right]=\frac{\hbar^2}{12 m}
\frac{1}{\rho^{5/3}(r)}\frac{\partial}{\partial r} \nabla^2 \rho(r)\;.
\label{leili}
\end{equation}
One can reduce  the order of the differential equation~(\ref{leili})
by explicit integration and  by using the ``boundary condition''
$[t'(r)/\rho'(r)]_{r=0}=(M+2)\hbar 
\omega$ from Eq.~(\ref{turning}) to obtain 
\begin{equation}
t'(r)= \frac{(M+2)\hbar \omega}{\rho^{2/3}(0)}
\rho^{2/3}(r)\rho'(r)+\frac{\hbar^2}{12m} \rho^{2/3}(r)\rho'(r)
\int_0^r ds\,\frac{1}{\rho^{5/3}(s)}\frac{\partial }{\partial
s}\nabla^2 \rho(s)\;.
\end{equation}
A second integration leads to the final result 
\begin{equation}
t(r)=\frac{t_W(r)}{3}+\left\{C+\frac{\hbar^2}{3m}\int_0^r
ds\frac{[\rho'(s)]^2}{\rho^{8/3}(s)}
\left[\frac{1}{2s}+\frac{\rho'(s)}{3\rho(s)}\right]\right\}\rho^{5/3}(r)   
\label{func}
\end{equation}
where 
$t_W(r)=(\hbar^2/8 m)[\rho'(r)]^2/\rho(r)$ is the von Weizs\"acker
``surface''  
contribution  to the  kinetic energy
density \cite{weis} and the constant $C$ is given by 
$C=3(M+2)\hbar\omega/(5\rho^{2/3}(0))-(\hbar^2/20m)(\nabla^2
\rho(r))_{r=0}/\rho^{5/3}(0)$. We have assumed $\partial
\rho(r)/\partial r|_{r=0}=0$.

The form $\rho^{5/3}(r)$ in Eq.~(\ref{func}) suggests, with
of course the exact density $\rho(r)$, the introduction of the
Thomas-Fermi (TF) kinetic energy density defined by 
\begin{equation}
t_{TF}(r)=c_k \rho^{5/3}(r)\;.
\end{equation}
 The constant $c_k$ is a multiplicative factor
which we do 
not need to specify for the present purposes.
Thus, as in earlier work in this Journal
\cite{vandoren} on the one-dimensional case of harmonic confinement,
we obtain that $t_{TF}(r)$ and $t_W(r)$ enter as building blocks 
the non-local kinetic energy density functional~(\ref{func}).

\section{Summary}

In summary, we have derived  two relatively simple equations,
namely Eqs.~(\ref{vir3d}) 
and~(\ref{turning}), the first being a differential form of the virial
theorem and the second
relating turning points of kinetic energy density $t(r)$ and particle
density $\rho(r)$. These  allow {\it (i)} the differential
equation~(\ref{eqdiff}) for $\rho(r)$ and {\it (ii)} the non-local
kinetic energy density $t(r)$ in Eq.~(\ref{func}) to be obtained.

\section*{Acknowledgements}
This work was partially supported by MURST through
PRIN2000. N.H.M. wishes to acknowledge
generous support from SNS for a stay in Pisa during which much of his
contribution to the present work was carried out. N.H.M. also
acknowledges valuable discussions with Dr. T.~G\'al and with
Professors L.~C. Balb\'as, A. Holas and \'A. Nagy on the general area
embraced by the present investigation.

\end{document}